\documentclass[review]{elsarticle}

\usepackage{hyperref}
\usepackage{bm}
\usepackage{booktabs}

\usepackage{url}
\usepackage{booktabs}
\usepackage{subfig}
\usepackage{graphicx}
\usepackage{float}
\usepackage[cmex10]{amsmath}
\usepackage{multirow}
\usepackage{amssymb}

\begin{document}

\begin{frontmatter}

\title{\normalsize{Homophily of Music Listening in Online Social Networks}}

\author{Zhenkun Zhou, Ke Xu}
\address{State Key Lab of Software Development Environment, Beihang University, Beijing, China}
\author{Jichang Zhao$^*$}
\address{School of Economics and Management, Beihang University, Beijing, China\\
$^*$Corresponding author: jichang@buaa.edu.cn
}

\begin{abstract}

Homophily, ranging from demographics to sentiments, breeds connections in social networks, either offline or online. However, with the prosperous growth of music streaming service, whether homophily exists in online music listening remains unclear. In this study, two online social networks of a same group of active users are established respectively in Netease Music and Weibo. Through presented multiple similarity measures, it is evidently demonstrated that homophily does exist in music listening of both online social networks. The unexpected music similarity in Weibo also implies that knowledge from generic social networks can be confidently transfered to domain-oriented networks for context enrichment and algorithm enhancement. Comprehensive factors that might function in formation of homophily are further probed and many interesting patterns are profoundly revealed. It is found that female friends are more homogeneous in music listening and positive and energetic songs significantly pull users close. Our methodology and findings would shed lights on realistic applications in online music services.

\end{abstract}


\end{frontmatter}

\section{Introduction}
\label{sec:intro}


One of the best established findings in social networks is that people who are friends exhibit plenty of similarities in human behaviors~\cite{Mcpherson2001Birds,de2010similarity}. Friendship relations, either offline or online, in which individuals socially interact, involve a need for shared mutual understandings. Tremendous efforts have been devoted on homophily of social networks from many aspects, ranging from demographics~\cite{chmiel2011negative} to mental states~\cite{bollen2011happiness,fan2014anger}. Even it is revealed from the recent study that personality similarity exists among close relationships~\cite{youyou2017birds}. Particularly, with the booming of online social media, previous studies have extensively investigated the homophily of behaviors in online social networks~\cite{brown2007word,mislove2010you,centola2010spread} and the friend similarity that exists in face-to-face offline networks is identically revealed. However, as a prominent element of daily life that possesses the cultural universality~\cite{blacking1995music,North2004Uses,wright2013listening}, music listening is rarely explored in the context of social networks and little knowledge is established about the behavior referring to homophily, especially in the circumstance of online social networks.

In fact, individuals that embedded in social networks come across music of varying categories, including vast kinds of genres, languages and moods, and continually judge whether or not like the music~\cite{frith2002music}. In addition, music is always shared with families, friends and other folks socially around us. Before times of the Internet, CDs and cassettes were the main media to record the music and the music communication was thus limited and awkward~\cite{bostrom1999mobile}. Yet in last twenty years, with the prosperous development of the Internet, portable music players have exploded in popularity, promoting the music communication essentially~\cite{Holmquist2005Ubiquitous,Rondeau2005For}. Friends are willing to exchange the iPods from each other. And now, the music streaming platforms offer low-latency access to the large-scale database, such as Spotify~\cite{kreitz2010spotify}, Last.fm~\cite{henning2008mendeley}, QQ Music~\cite{priest2006future} and Netease Music~\cite{fung2007emerging}. Since then, people exchange the music with each other online and share the amazing music they respectively like, and music even creates interpersonal bonds between different individuals in turn~\cite{boer2011shared}. Though evidence of music similarity in offline friendships has been demonstrated~\cite{Selfhout2009The}, the relationship between listening similarity and online social networks has not been comprehensively explored and understood yet. Specifically, questions like whether we enjoy the music at which the online friends are enchanted, in other words, whether the homophily of music listening exists in online social networks, still deserve a systemic investigation.


Until recently, empirical research willing to answer the questions about music listening had to depend on interviews and surveys in controlled environments~\cite{sloboda1999everyday,greenberg2016song} with inevitable limitations in both data scale and granularity. While in fact, the technological and societal evolutions that sustain the emergence of online music listening indeed provide unparalleled opportunities for human behavior understanding. Detailed footprints, including where, when and how massive individuals listen music can be regarded as a big-data window and through which the homophily in music listening can be collectively or individually probed and studied thoroughly. For instance, Netease Music, one of the most popular online music providers in China, contributes the high-quality music streaming service to millions of users and accordingly accumulate the detailed behavior records of these users continuously. According to the official report of Netease Music, its active users is over 200 million. With music playlists creation being the core listening pattern, each day users establish around 420 thousand playlists and user-generated playlists total 800 million. In the half of 2016, users play songs 1.82 billion times and the duration amounts to 7.2 billion minutes, implying the impressive vitality of users in online music listening. Even more inspiring, Netease Music develops one extraordinary trait of socializing its users. Specifically, it firstly provides a domain-specialized social network through which users engage in sharing interests of music. Like generic online social networks, users can be networked through following others, not only ordinary users but also artists. Indeed, the social network sourced in music listening profoundly facilitates the acquiring and sharing of music interests, implying an ideal entanglement between social ties and music listening for the present study. Thus anonymous digital traces of massive users are collected to quantitatively support the investigation of homophily in music listening.


However, domain-oriented social networks like the one established by Netease Music can not be a typical representation of online social networks that are generally resulted by comprehensive factors and music might be just one of them. Specifically, Netease music social network relies predominately on the musical interest and its digital traces are insufficient to describe other individual traits. In the meantime, evidence of musical preferences being linked to individual traits like personalities~\cite{greenberg2016song}, cognitive styles~\cite{greenberg2015musical} and even socioeconomic statuses~\cite{park2015understanding} has been extensively demonstrated, implying the consideration more generic online social networks. We argue that aiming at a systemic understanding, it is necessary to study listening practices based on other more typical social networks in which users are jointed sophisticatedly but realistically by psychological traits, extensive interests or other individual characteristics. Considering the booming of social media in recent decades, prosperous networks like Twitter or its variant Weibo of China that aggressively replicate offline social networks to online counterparts, can be ideal targets. These online social networks are natural, long-term and diverse and thus the objective footprints of massive individuals can be promising proxies for the present study. Nonetheless, it is still extremely difficult to correctly match each individual of the music social network to the identical one in networks like Weibo, which is the essence of embedding music listening into a generic online social network. Very fortunately, users can log into Netease Music through their Weibo accounts and along this line we can obtain the digital traces of Weibo for these users. Therefore, it is possible to further study the similarity of music listening for friends that linked in Weibo. In addition, demographics and tweets in Weibo are excellent supplements for enriching individual characteristics. We can even explore which key factors influence the similarity of music listening.


Starting from the above motivations and assumptions, in this study, digital footprints of over seventy thousand individuals from both Netease Music and Weibo are thoroughly collected through a novel crawler. Then 25, 953 active users with plenty of times for listening and tweeting are sampled as the subjectives for further explorations. Two online social networks, the Netease network and Weibo network are respectively constructed through user followings in Netease Music and Weibo. To examine whether the music listening is homogeneous for users jointed by online social networks, similarities from six perspectives are defined and measured. In order to investigate the crucial factors influencing the music listening homophily, subjectives are clustered into different categories from multiple perspectives like social attributes, musical preferences or etc. Our results demonstrate that friends linked by online social networks indeed appreciate the identical songs and possess the similar music preferences. As for gender, patterns for music listening between female friends are closer than male friends, implying that women are more sensitive for emotional expression through music~\cite{wells1991emotional,robazza1994emotional}. Listening practices of the friends with common music preferences in the current culture are similar especially for music languages (Chinese) and genres (pop and folk). With regard to musical mood, the users who enjoy exciting, wild and happy music share more similarities in music listening. It is also difficult for users owning high musical diversity to find friends with similar musical tastes. The present study confirms evidently the existing of homophily in music listening of online social networks and elaborately clarify the roles of human demographics and music traits in influencing the homophily. Our findings would shed insightful lights on music recommendation and friend suggestion in online applications. We merge different social circles of an individual and surprisingly reveal that generic social networks like Weibo still demonstrate significantly the homophily that intuitively only exists in domain-oriented networks. It indeed implies that rich information in general social media can be confidently introduced into the study of specialized social networks~\cite{carmagnola2009user}.

\section{Methods}

\subsection{Netease Music Dataset}
\label{netease}

Digital service providers began to amass large user bases, offering the primary sources of digital music streaming via the Internet increasingly. New innovations, including digitalization and the Internet, transform the existing landscape over the past decade and attracted new artists and listeners into the fields, and digital music streaming services has also profoundly reshaped the user behaviors. China follows this global trend and has become a leading digital country in terms of music services. The Netease Music is one of the most trending music streaming providers in China, whose special trait is the music social network. Users are allowed to follow others in the platform, such as friends and artists. Therefore, this music platform has become a new and booming social network of domain-oriented.

The web site of Netease Music provides abundant online information about users, playlists and music. We develop an excellent crawler aiming at Netease Music platform first to perform the data collection. Our crawler adapts massive agent servers and each agent will simulate a real user to visit pages and click links from the browser. By traversing the following links, from August in 2016 to May in 2017, we obtain a dataset of over 200 thousand users, 1.5 million playlists and 3.2 million songs.

The collected data referring to users include the following lists, historical listening and like records. For each user, the following list contains the complete ID numbers of users (UID) that she/he followed. The top 100 songs listened by one user are recorded in her/his historical listening. The platform provides the `like' button to label their favorite playlists or songs which are restored in like records. According to the historical listening and like records, we obtain the corresponding detailed playlists and attributes of songs. In each playlist, ID numbers of songs (SID) it contains and style tags labeled by its creator are both collected. The attributes of a song include the album, artists and several acoustic features.

\subsection{Weibo Dataset}
\label{weibo}

Tremendous social relationships are forged in popular online services like Facebook, Twitter and Weibo with the explosive development of online social networks. Twitter, as the most popular social networking and micro-blog service, enables registered users to read and post short messages, so-called tweets. At the beginning of 2016, Twitter had reached 310 monthly active users (MAU). As of the third quarter of 2017, it averaged at 330 million MAU. However, more people are using Weibo, the Chinese variant of Twitter now. According to the Chinese company's first quarter reports, it has 340 million MAU, 30\% up on the previous year, implying that social relationships and online behaviors of massive individuals from Weibo can be sensed and profiled with unparalleled richness and granularity. Moreover, replicating offline social ties to online and replacing face-to-face communication with interactions of cyberspace have essentially digitized the daily life and reshaped the social networks, suggesting that the online social network is of fundamental significance for explorations in both social theories and applications.

In the fore part mentioning Netease Music Dataset, 74,056 users connect their Weibo accounts to Netease Music, offering us an opportunity to establish a perfect match between these two different platforms. Employing the crawler agents, their corresponding ID numbers in Weibo are obtained. With the help of Weibo's open APIs, for each ID, the public profile, following relationships and historical tweets can be accordingly collected. And in order to speed up the the collection, cloud servers are also extensively utilized. The profiles of Weibo users include demographics (e.g. gender), social attributes (e.g. the number of followers) and other individual information. For each Weibo user, the following relations contains the Weibo ID numbers (WID) of users it follows. Unfortunately, due to the official limit on viewing following list, we only obtain the latest 1,000 following relations. Nevertheless, according to theory of Dunbar's number, the maximum size of ego-network averages 150~\cite{DUNBAR,hill2003social,zhao-dunbar}, implying that the latest one thousand followings could sufficiently reflect recently active social connections the user possesses. Similarly, due the limit on tweets viewing, for each Weibo user only its latest 1000 (or less if not enough) tweets can be obtained.

\subsection{Two Social Networks}
\label{sec:sn}

\begin{figure}
\centering
\subfloat[]{\includegraphics[width=10cm]{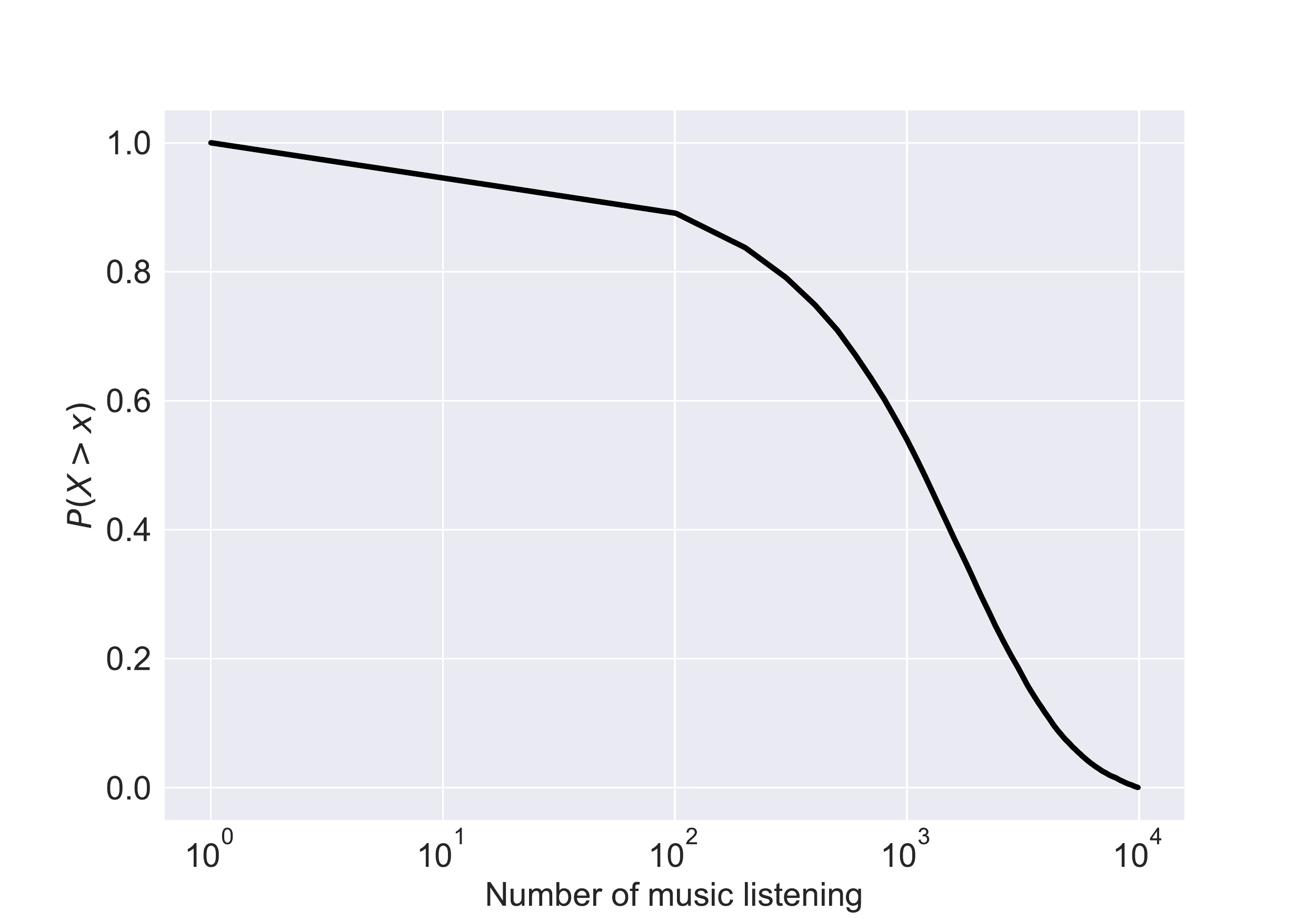}
\label{fig:dist_listen}}
\hfil
\subfloat[]{\includegraphics[width=10cm]{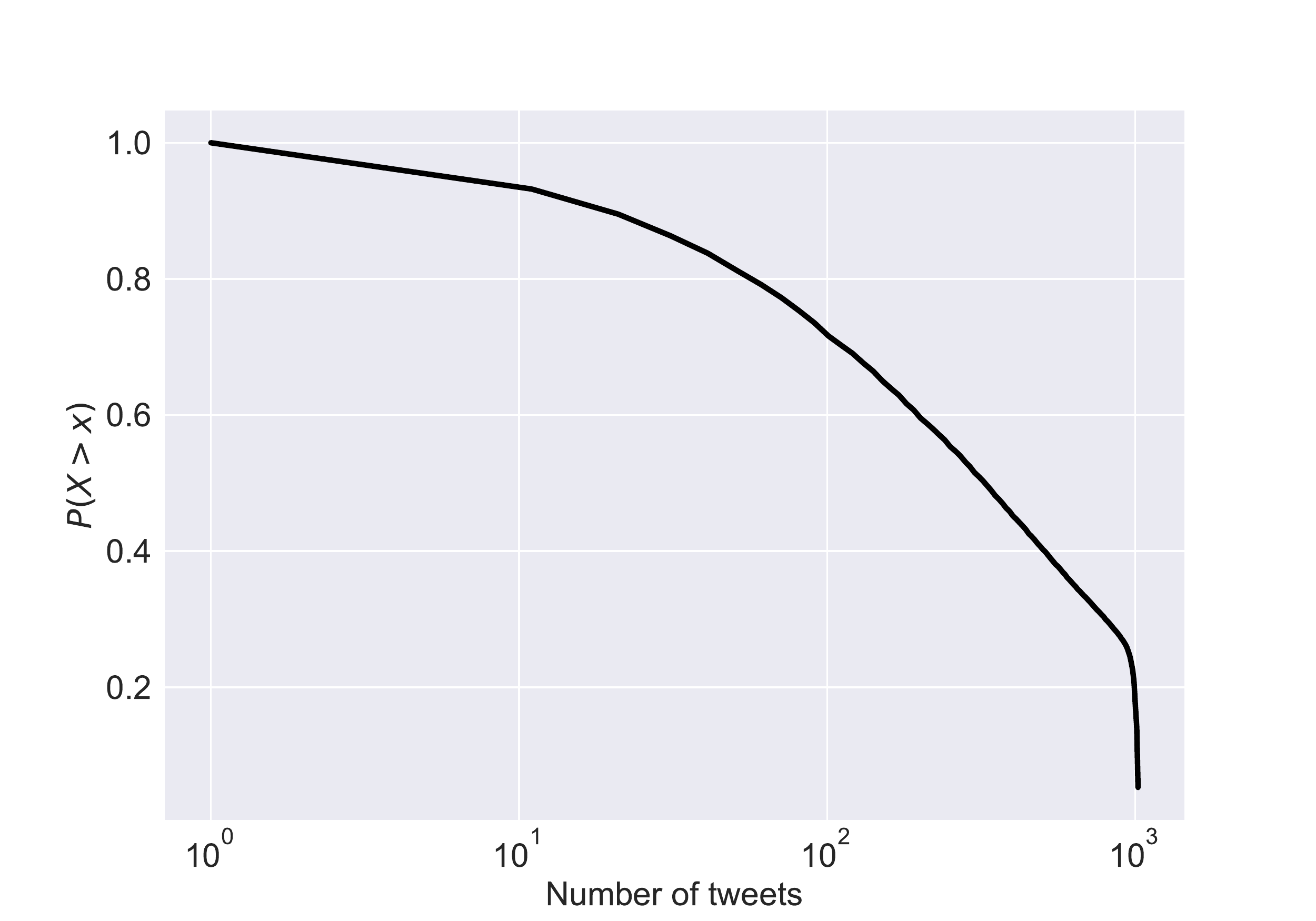}
\label{fig:dist_tweet}}
\hfil
\caption{CCDF of the number of music listening and tweeting.}
\label{fig:number}
\end{figure}

We further refine the subjects of the present study by selecting active users from both datasets. Specifically, active users are defined as those have listened complete songs at least 1000 times and posted over 100 tweets. The complementary cumulative distribution function (CCDF) of the number of users in music listening and tweeting is depicted in Fig~\ref{fig:number}. As can be seen in Fig~\ref{fig:dist_listen}, for the majority of users, the number of music listening is lower than $10^4$. The percent of users that listen music over 1000 times is lower than 55\%. As shown in Fig~\ref{fig:dist_tweet}, about 70\% of users tweet 100 times at least. Note that owing to the limit of Weibo, the maximum of number of tweets is 1000. After the refinement, totally 25,953 active users are filtered out to be subjectives of further explorations.

\begin{figure}
\centering
\subfloat[]{\includegraphics[width=10cm]{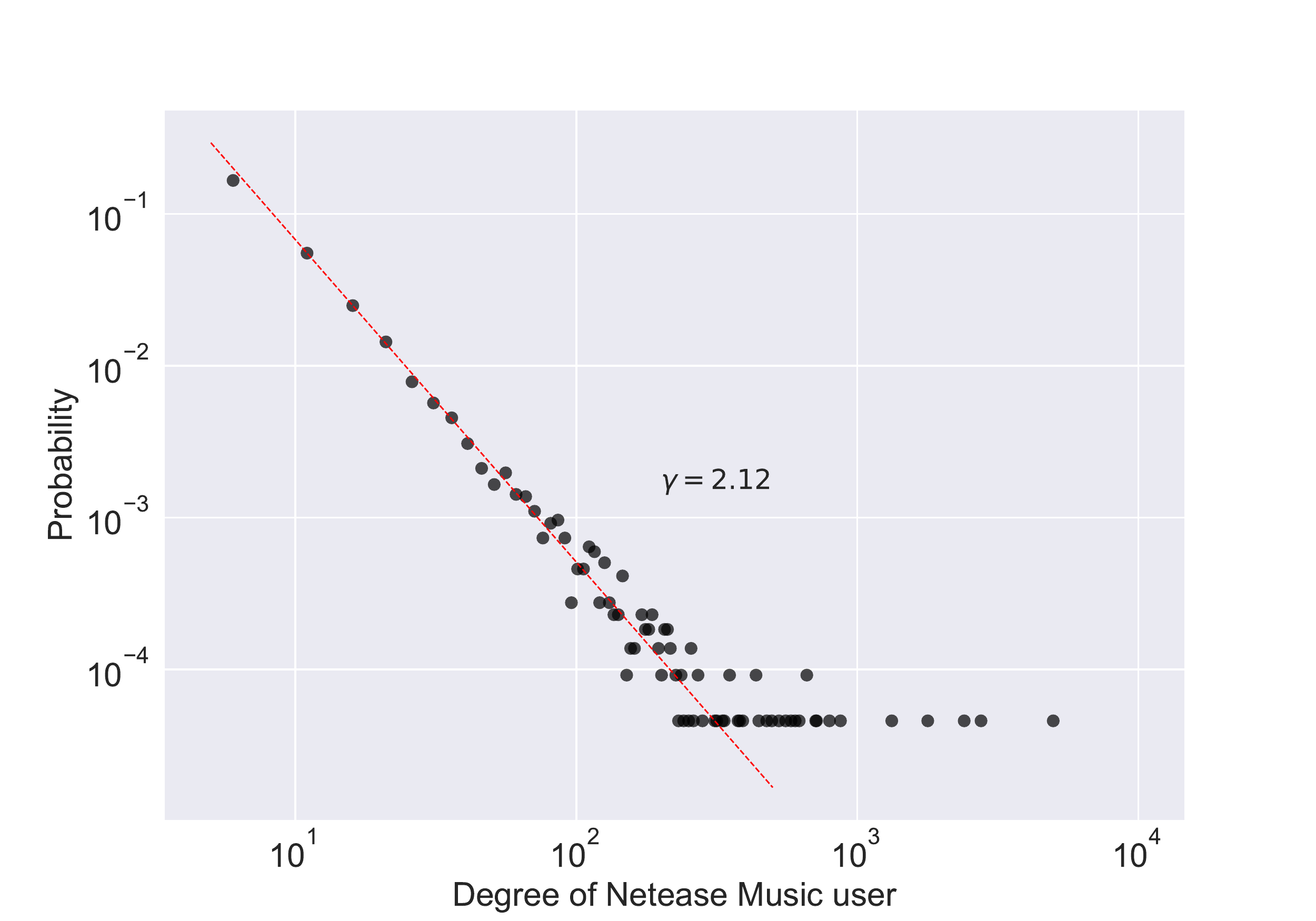}
\label{fig:dist_netease}}
\hfil
\subfloat[]{\includegraphics[width=10cm]{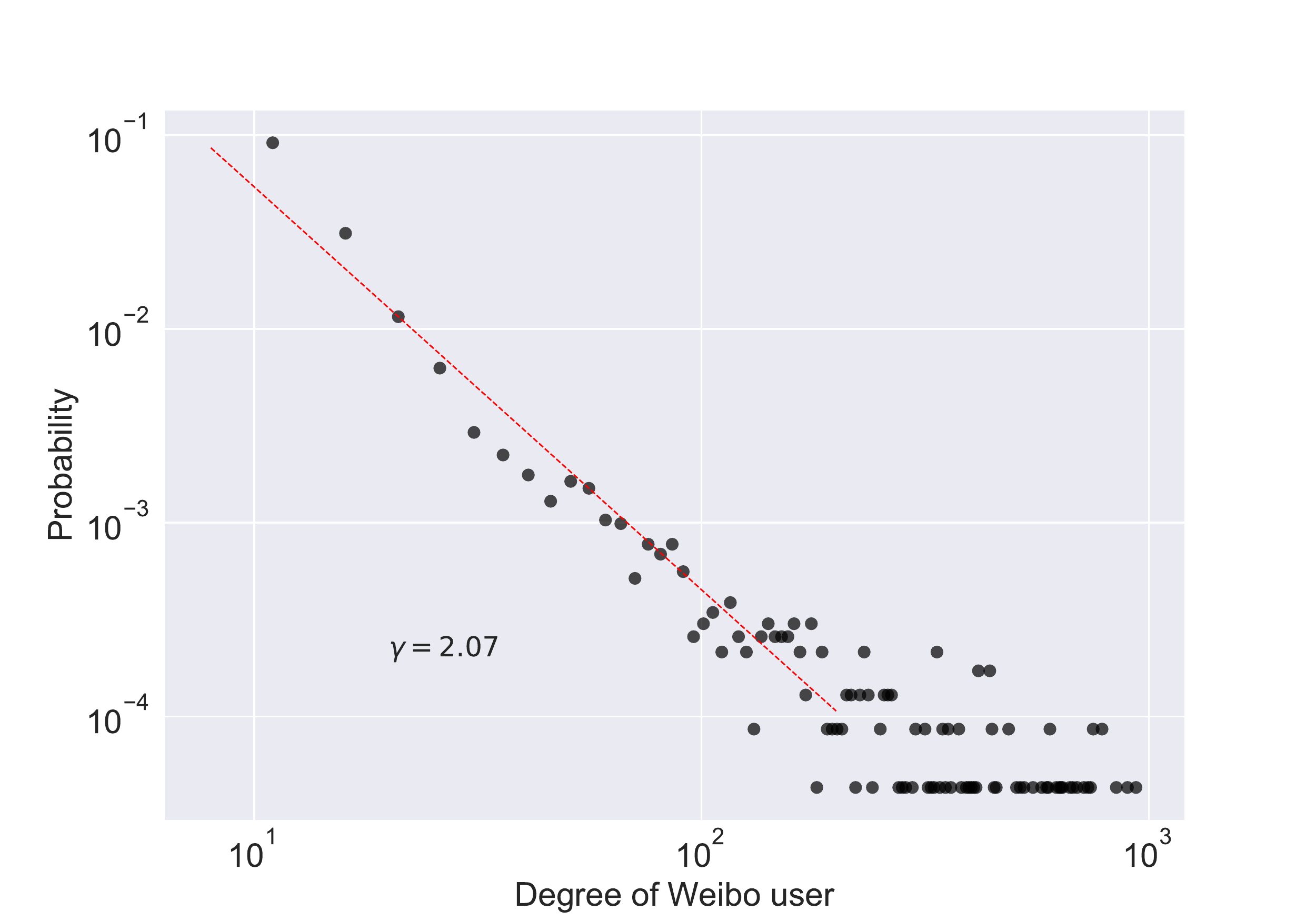}
\label{fig:dist_weibo}}
\hfil
\caption{Probability distribution of the degree of Weibo network and Netease Music network. The degree exponent $\gamma$ of networks are respectively 2.12 and 2.07 ($2 < \gamma < 3$).}
\label{fig:degree}
\end{figure}

Regarding to the same group of active users, two social networks are then respectively established from the following relationships in Netease Music and Weibo. The first one, named Netease network, contains 89,988 links. The second one, named Weibo network, contains 112,753 links. And these two networks share the 25,953 identical nodes, implying that we construct an ideal scenario to reveal the homophily of music listening in both domain-oriented social network and generic social network. It is worthy noting that here we regard social networks as undirected graphs. The degree distributions of Netease network and Weibo network are demonstrated in Fig~\ref{fig:degree} and the power-law trending implies the consistence to existing understanding of typical social networks~\cite{newman2010networks}. As for Netease network ($\gamma=2.12$), the distribution evidently shows a long tail, indicating there are few users with large amount of relationships in social network of music-oriented. As for Weibo network ($\gamma=2.07$), a power-law like distribution is observed as degree lower than 500, however, a exponential cutoff then emerges and the maximum degree is 100, both due to the API limit of Weibo.

\subsection{Similarities of music listening}
\label{similarity}

Based on the personal historical listening and like records, we define six types of similarities with respect to music listening for any pair of users $A$ and $B$ in both social networks. And higher similarities suggest more significant homophily in social networks.

The historical listening records for $A$ and $B$ are denoted as $\bm{H}_{A}$ and $\bm{H}_{B}$, respectively and in which top 100 songs of high-frequency are included. The historical similarity $sim_{song}$, indicating the number of co-occurrence songs in top records can be thus defined as
\begin{equation}
sim_{song} = \ \mid\bm{H}_{A} \cap \bm{H}_{B} \mid \ .
\end{equation}

For presenting measures of similarity from aspects of music traits, we first develop ways to label songs and users in terms of traits vectors. Given the missing of tags of songs in Netease dataset, the tags of playlists can be fused to infer traits vectors of songs. Specifically, Netease officially provides 6 tags of languages, 24 tags of genres, 13 tags of moods and 12 tags of scenes and among these 55 tags at most three are selected to label a playlist by its creator. Assuming that a song can be labeled confidently by tags of the playlists it belongs to, a 55-dimension traits vector (initialized to be zero) can be accordingly derived for each song by examining 1.5 million playlists, i.e., a tag occurrence will result in addition of 1 to the corresponding dimension. However, considering the fact that values of the traits vector of a song can be significantly influenced by the number of playlists it appears, we split the long vector into four sub-vectors including $S_{language}$, $S_{genre}$, $S_{mood}$ and $S_{scene}$ ($S$ refers to $\bm s$ongs) and then do normalization on each sub-vector to avoid bias. To be specific, we calculate the proportion of each tag respectively in four vectors and the sum of values in each sub-vector should be $1$. In fact, $S_{language}$, $S_{genre}$, $S_{mood}$ and $S_{scene}$ can be features to appropriately represent the category distribution of songs. Following the same idea, supposing that music preferences of a user can be well reflected by the top songs listened, the traits vectors of users can be inferred through
\begin{equation}
U_{c} = \sum_{i=1}^{100}S_{c,i},
\end{equation}
in which $c$ stands for traits and can be language, genre, mood or scene. Finally, based on the feature vectors of 3.2 million songs, traits vectors for 25,953 active users are derived and normalized to z-cores. Then the similarity of user pair $A$ and $B$ can be intuitively measured through cosine distance between their traits vectors, which is defined as
\begin{equation}
sim_{c} = \frac{U_{c, A} \cdot U_{c, B}}{\mid U_{c, A} \mid \mid U_{c, B} \mid}.
\end{equation}
Note that $-1 \leqslant sim_{c} \leqslant 1$ and values closer to 1 indicate more similar music preferences. And as $c$ varies over different music traits, we accordingly obtain four music preference similarities from perspectives of languages, genres, moods and scenes.

In addition, the similarity of favorite songs can also reflect the strength of homophily in music preference. Different from historical listening records, the number of songs in like record is not limited. Therefore we define the Jaccard distance as the similarity for like records between users $A$ and $B$ as
\begin{equation}
sim_{like} = \ \frac{\mid\bm{L}_{A} \cap \bm{L}_{B} \mid}{\mid\bm{L}_{A} \cup \bm{L}_{B} \mid}
\end{equation}
in which $\bm{L}$ stands for the set of one's favorite songs.

To sum up, from the above we obtain six similarities referring to music listening and higher similarities indicate more significant homophily in music listening of online social networks. These measures pave the way for quantitative investigations of homophily in this study.

\subsection{User classifications}
\label{sec:user_cla}

Many factors in social networks might influence the homophily between friends. In order to reveal the crucial factors, demographics, social attributes and listening behaviors of active users are further probed for detailed understanding of their roles in homophily of music listening. We argue that regarding these factors as features, active users can be thus clustered into groups from multiple perspectives and discriminations of inter-groups offer windows of homophily investigation and factor weighing.

Gender is one of the most significant demographics in understanding human behaviors. We extract gender from Weibo profiles and split active users into 10,388 female ones and 15,565 male ones. Whether a user is officially verified by Weibo can be an indicator of influence and through the `verified' labels active users are split into 24,368 non-verified ones and 1,585 verified ones. In addition, social attributes that reflect users' ranks in the social network can be well modeled through the number of the follow\textbf{er}s ($NER$) as well as $RFF$ that defined as the \textbf{r}ate of \textbf{f}ollowing numbers to \textbf{f}ollowed numbers ($log ( \frac{NER+1}{NEE+1} )$, in which $NEE$ refers to the number of the users' followees). Given the fact that both $NER$ and $RFF$ are continuous variables, the discretization based on $K$-means clustering is employed to categorize active users into groups of \textbf{low} rank and \textbf{high} rank respectively based on these two attributes.

Patterns of online music consumption could be sufficiently reflected in terms of users' music preferences that result in behavioral differences directly. Hence active users can also be clustered into groups from the perspective of music preference. The traits vector $U_{c}$ can be features representing users in clustering and the preference feature matrix of $c$ is accordingly constructed for all users. The matrix is de-correlated by principal component analysis (PCA) with full covariance and then the approach of $K$-means is employed to cluster active users into groups. Clusterings based on all traits are denoted separately as $C_{mood}$, $C_{language}$ and $C_{genre}$, indicating that for each trait active users are grouped into three clusters with divergent patterns of music preferences. In order to better interpret the clusterings, for groups of each clustering, the group feature defined as the mean user feature within the group is calculated. Then the semantic of each group is explained by the top 3 attributes of the group feature. Table~\ref{tab:clas} reports the clusterings and group interpretations. As can be seen, groups of each clustering could be well explained by the top attributes, suggesting that our methods indeed capture the music preferences and behavioral patterns in music listening can be effectively detected.

%
\begin{table}[]{}
\centering
\caption{User classifications from perspectives about music preferences.}
\label{tab:clas}
\begin{tabular}{@{}ccl@{}}
\toprule
Traits    & Groups & Interpretations                      \\ 
\midrule
         & 0        & Chinese (including Cantonese)                \\
language & 1        & Japanese, Korean                 \\
         & 2        & English, minority                \\ \midrule
         & 0        & rap, dance, alternative          \\
genre    & 1        & pop, folk                        \\
         & 2        & light, New age, classic,         \\ \midrule
         & 0        & exciting, wild, happy            \\
mood     & 1        & sad, missing, lonely             \\
         & 2        & fresh, sanative, easy            \\ \bottomrule
\end{tabular}
\end{table}

In the meantime, as exploited in previous efforts~\cite{van2001social,park2015understanding}, diversity of musical preferences is also an indicator of great significance in reflecting listening behaviors. We draw on the concept of entropy~\cite{alexander1996entropy} to define users' diversity in music preferences as
\begin{equation}
div = -\ \frac{1}{3} \sum_{c=1}^{3}{\sum_{i=1}^{n}p(u_{c,i}) log(p(u_{c,i}))},
\end{equation}
in which $p(u_{c, i})$ refers to the proportion of $i$-th attribute in user vector of trait $c$. Higher diversity implies the user's more appreciations for variety of experience in music listening. We then employ $K$-means to cluster active users into high-diversity and low-diversity groups respectively.

The above user classifications can be finally utilized to define the homophily of relationships in online social networks. According to eight groups of all the clusterings, these are totally 19 factors\footnote{The factors range from demographics to music traits and include $female$, $male$, $V0$, $V1$, $NER0$, $NER1$, $RFF0$, $RFF1$, $lang0$, $lang1$, $lang2$, $genre0$, $genre1$, $genre2$, $mood0$, $mood1$, $mood2$, $div0$ and $div1$.}​. As the dummy variable, the value of factor is 1 when the two linked users are classified to the same group, otherwise the value is 0. For example, Factors $female$ is 1 and $male$ is 0 when user $A$ (female) is linked with user $B$ (female). If $female$ and $male$ are both 0, indicating that $A$ and $B$ possess different genders.
\section{Results}
\label{sec:results}

\subsection{Existence of homophily}
Being direct indicators of homophily, similarities from different perspectives are first investigated in the two social networks constructed from Netease and Weibo. Meanwhile, in order to testify the significance of the similarity distribution, for each realistic network, in terms of shuffling nodes a random counterpart is also built for base lines.  

\begin{figure}
\centering
\includegraphics[width=10cm]{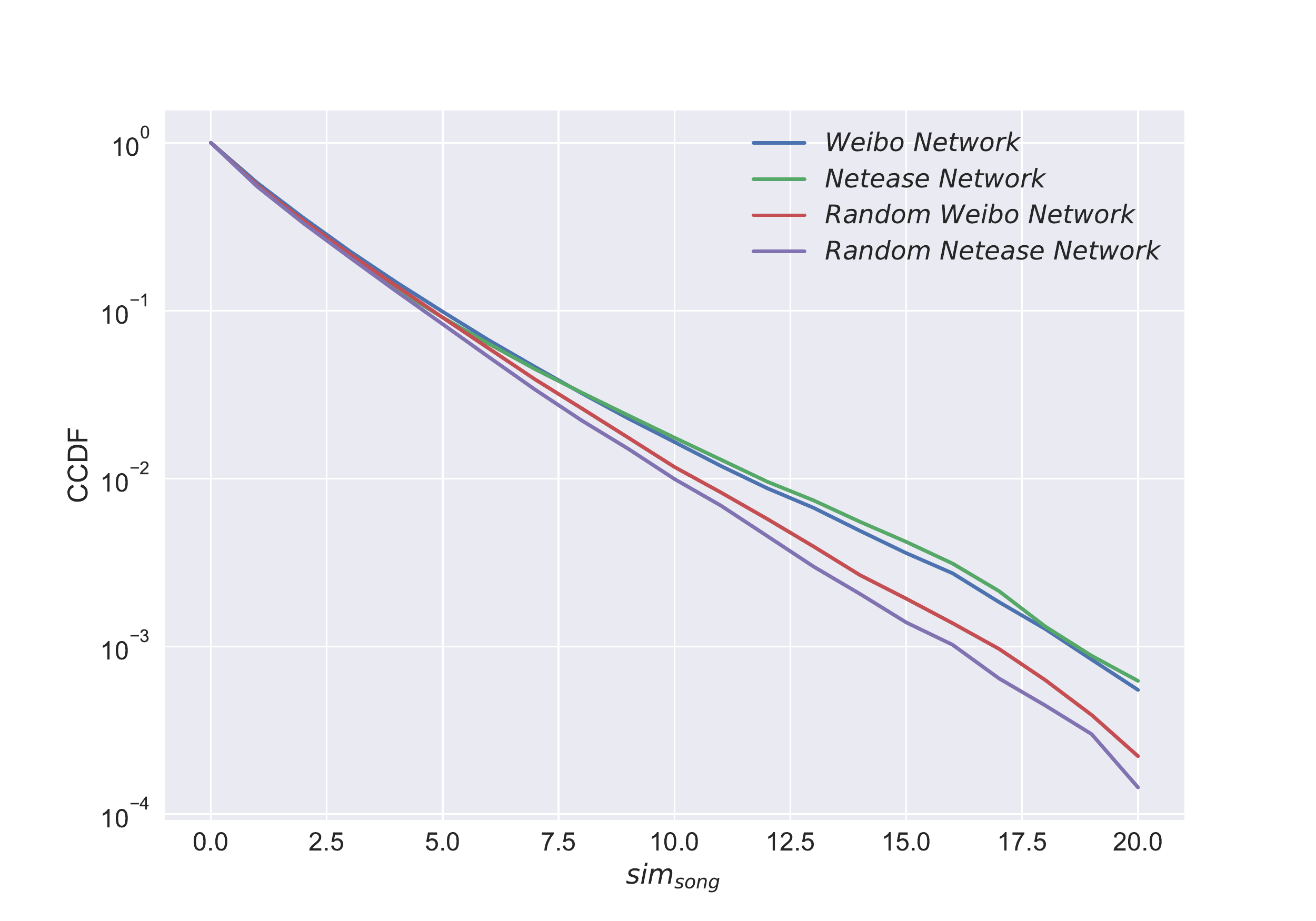}
\caption{Distribution of similarities of listening historical records. Through the two-sample Kolmogorov-Smirnov test, except for the two random networks ($p=0.99$), the distributions of other pairs have significant differences ($p<0.001, ^{***}$).}
\label{fig:sim_song}
\end{figure}

The CCDF of $sim_{song}$, as can be seen in Fig.~\ref{fig:sim_song}, demonstrates that similarities of actual networks are evidently higher than that of random networks, especially when $sim_{song} > 5$, indicating that friends in online social network inclines to listen same songs and the homophily in music listening significantly exists. Particularly, as compared to Weibo network, friendships in Netease network are slightly closer, implying that from the perspective of listening same songs, homophily is enhanced in the music-oriented social network.

\begin{figure}
\centering
\subfloat[]{\includegraphics[width=5cm]{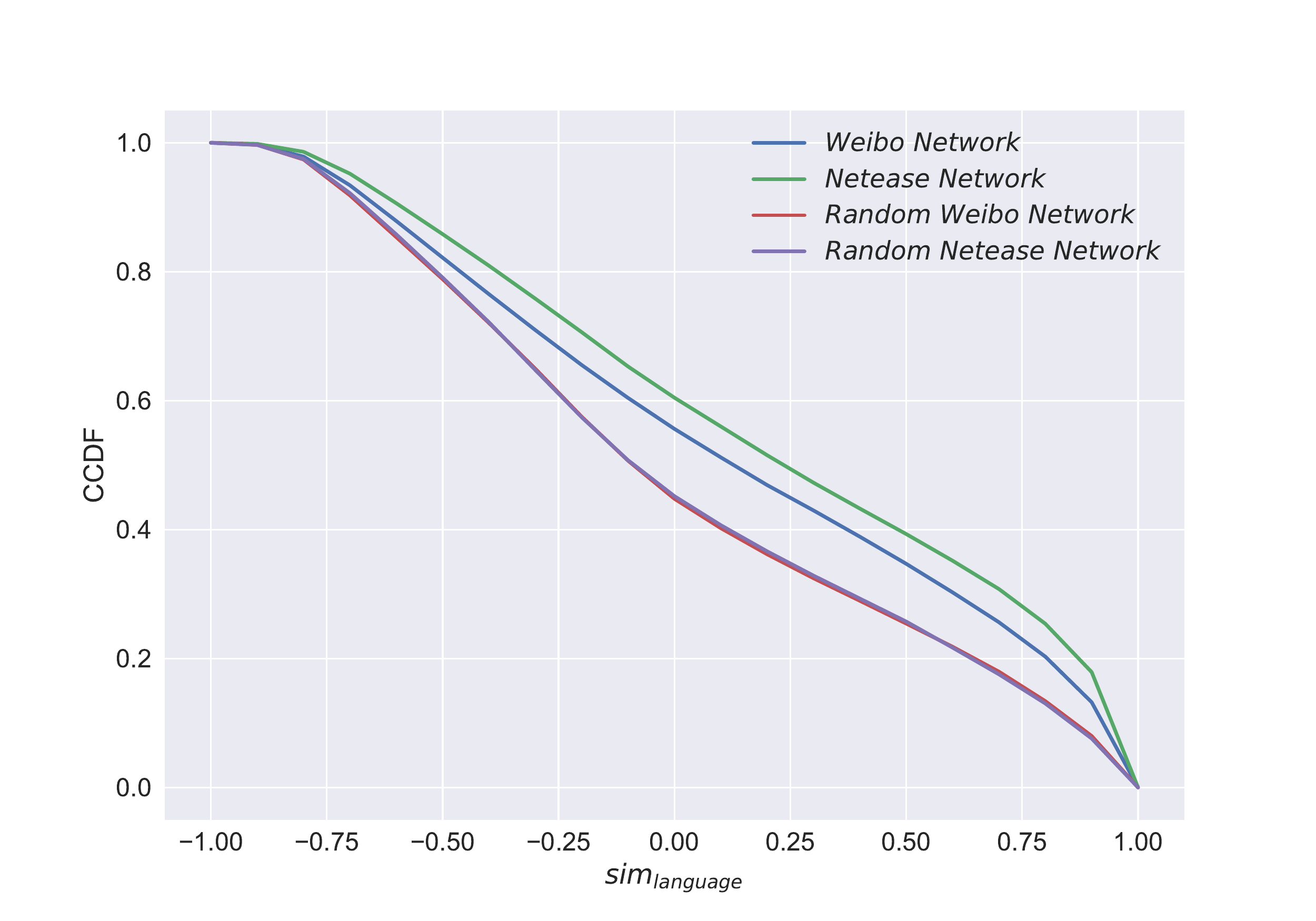}}
\hfil
\subfloat[]{\includegraphics[width=5cm]{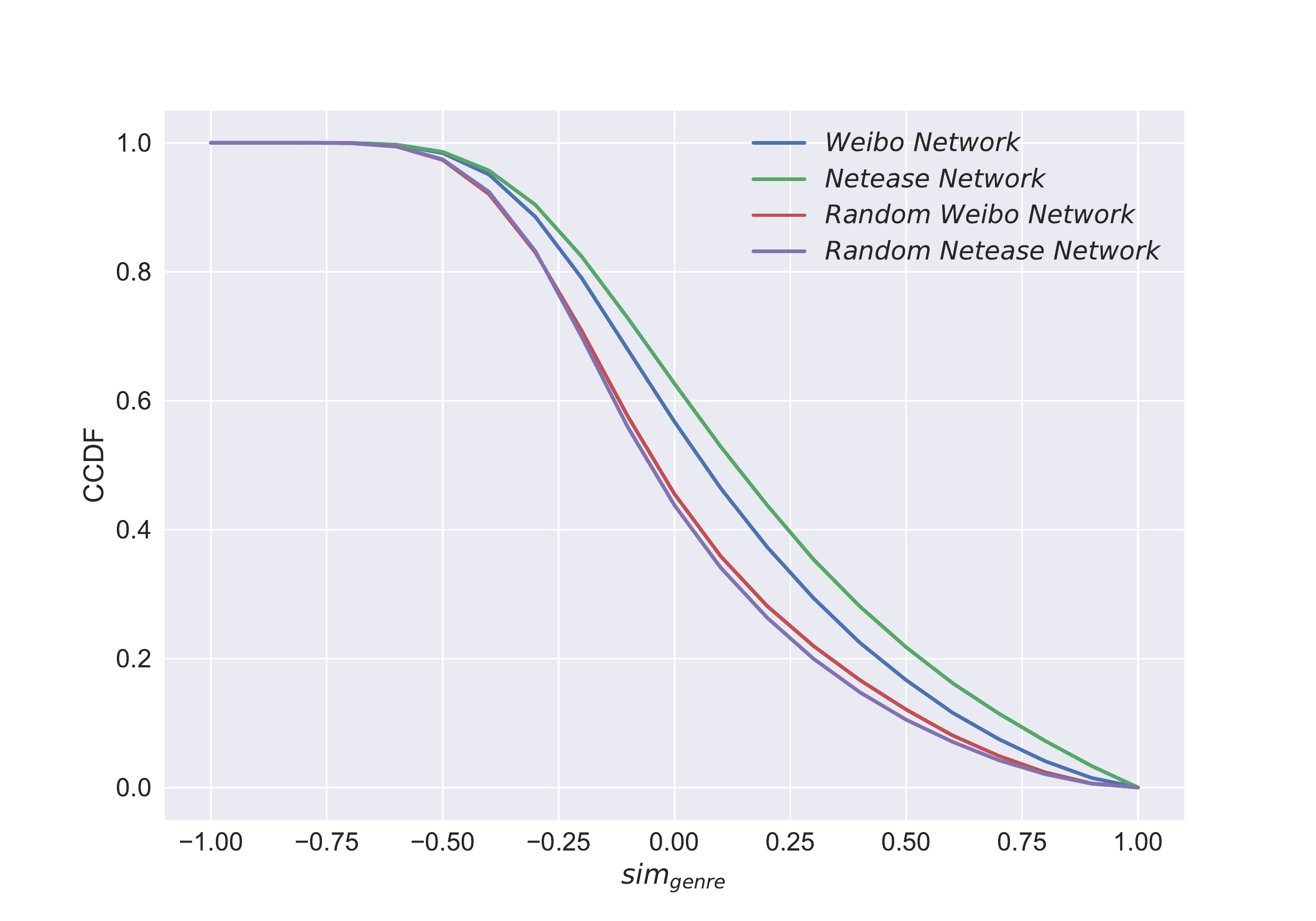}}
\hfil
\subfloat[]{\includegraphics[width=5cm]{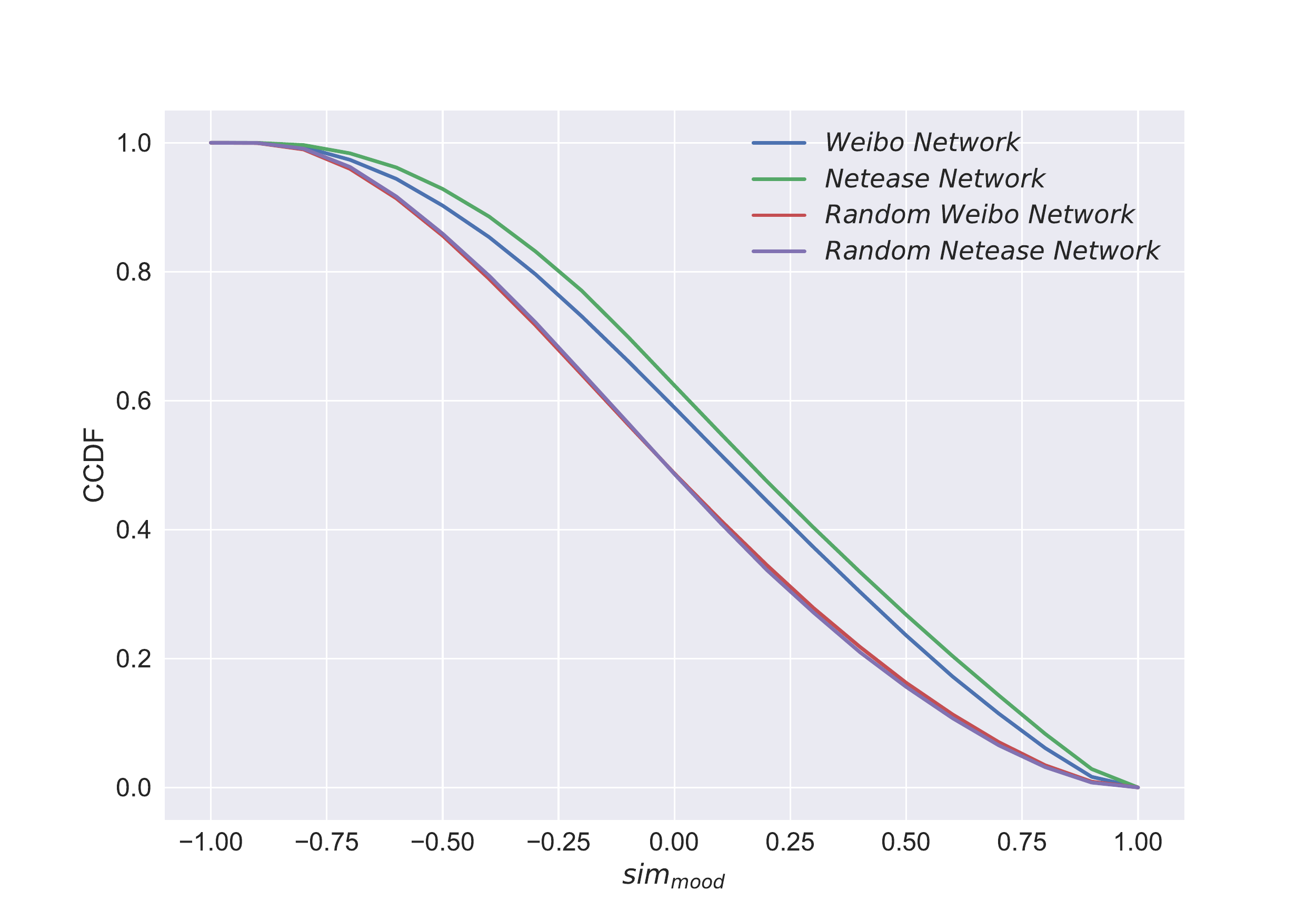}}
\hfil
\subfloat[]{\includegraphics[width=5cm]{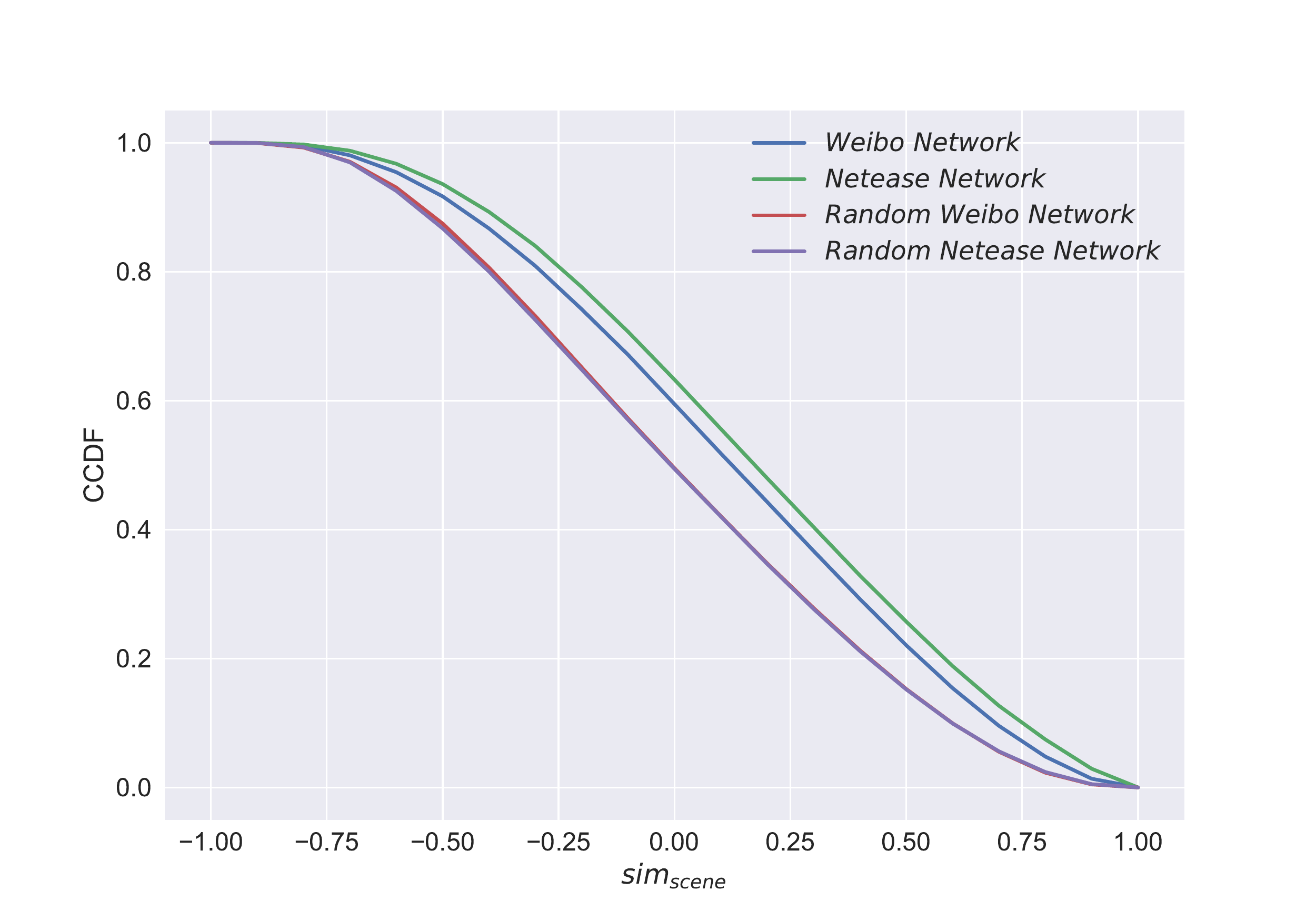}}
\hfil
\caption{Distribution of similarities of musical preferences. Through the two-sample Kolmogorov-Smirnov test, except for the two random networks ($p>0.05$), the distributions of other pairs have significant differences ($p<0.001, ^{***}$).}
\label{fig:sim_pre}
\end{figure}

\begin{figure}
\centering
\includegraphics[width=10cm]{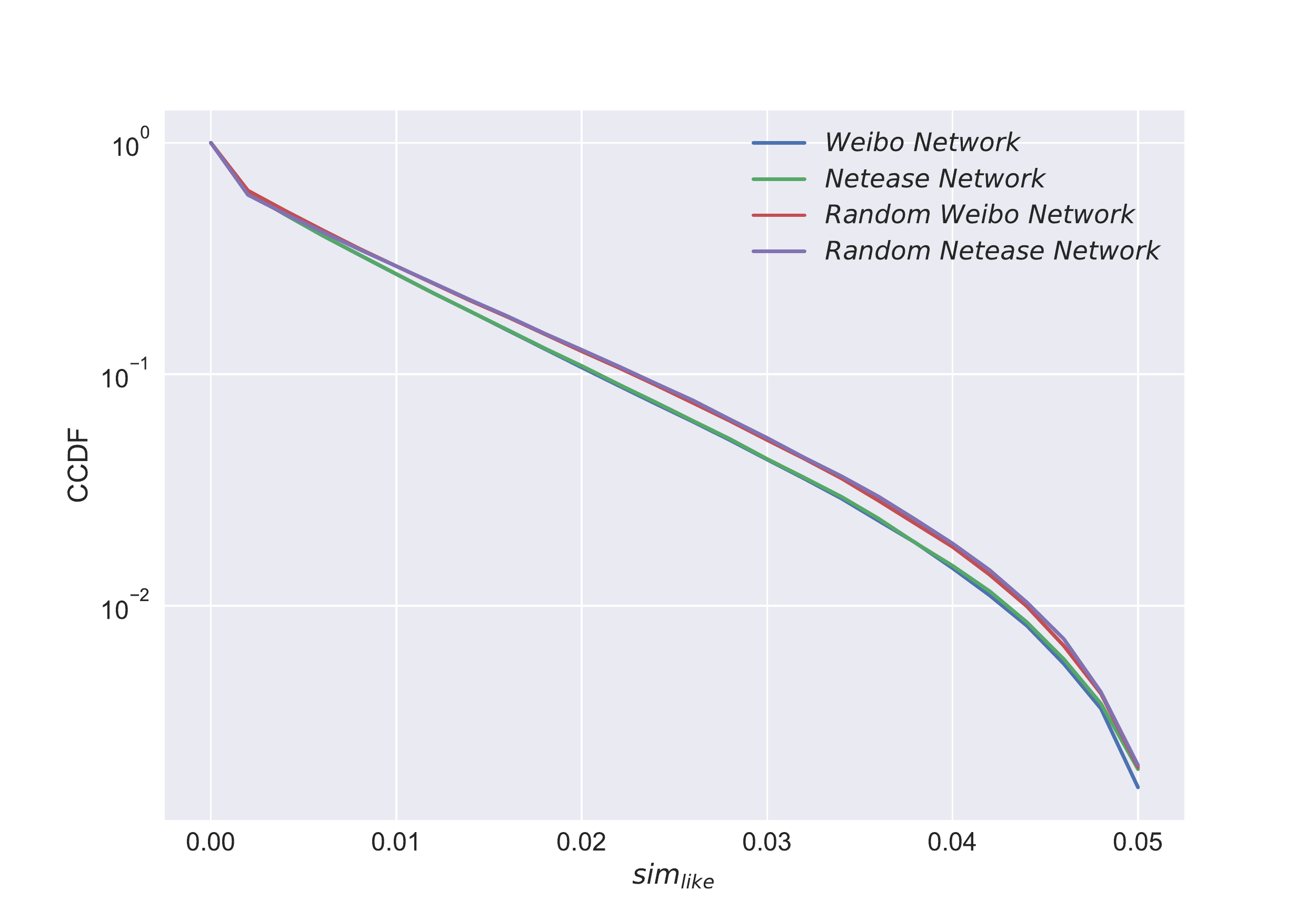}
\caption{Distribution of similarities of favorite songs between users. Through the two-sample Kolmogorov-Smirnov test, except for the two random networks ($p=0.163$), the distributions of other pairs have significant differences ($p<0.001, ^{***}$).}
\label{fig:sim_like}
\end{figure}

CCDFs of similarities in perspectives of music preferences are further investigated, as can be seen in Fig~\ref{fig:sim_pre}. Similar to observations of $sim_{song}$, it is consistently demonstrated that the friend similarity in music preferences is significant, suggesting the existing of homophily in music listening of online social network. Even more interesting, homophily in Netease network is more evident than that of Weibo network, implying again that friends are more homogeneous in domain-oriented social networks than the generic counterparts. However, as for the measure of $sim_{like}$ that shown in Fig~\ref{fig:sim_like}, there is no evidence for the existing of homophily because the similarity of friends in `like' behavior of online networks are close to or even lower than that in random networks.

To sum up, in terms of similarity metrics except the one regarding to `like' behavior, homophily of music listening in online social networks, especially the one of music-oriented, is significantly demonstrated and confirmed. Friends connected in online social networks indeed enjoy identical music and share similar music preferences. The absence of homophily in `like' behavior suggests that even in homogeneous online social networks, individual differences still exist across friends. Although the homophily in specialized network is slightly more significant, we indeed find the same behavior pattern in generic networks like Weibo where the relationships are bonded with more common interests. It further implies that social media like Weibo can be generic but typical instances of online social networks. We further explore the important factors which affect the listening similarity in the Weibo network.

\begin{table}[]
\centering
\caption{Results of multiple regression to predict $sim_{song}$ based on user classifications. The column of ``Coef.'' (coefficient) shows the influence of factors on homophily. Note: $^{**} p<.01, ^{***} p<.001.$}
\label{tab:multiple_regression}
\begin{tabular}{@{}llcccc@{}}
\toprule
Traits                     & Factors        & Coef. & Std. Error  & $t$-Value &    $Pr(>\mid t \mid)$\\

\hline \hline

\multirow{2}{*}{gender}    & female         & 0.369  & 0.021 & 17.25 &$^{***}$\\
                           & male           & 0.254  & 0.015  & 16.91  &$^{***}$\\ \midrule
\multirow{2}{*}{verified}  & V0             & 0.119  & 0.017   & 6.91 &$^{***}$\\
                           & V1             & 0.014         & 0.035   & 0.41 & \\ \midrule
\multirow{2}{*}{NER}       & NER0           & 0.444  & 0.042  & 10.58 &$^{***}$\\
                           & NER1           & -0.149  & 0.018  & -8.49 &$^{***}$\\ \midrule
\multirow{2}{*}{RFF}       & RFF0           & 0.517  & 0.031  & 16.72 &$^{***}$\\
                           & RFF1           & 0.036          & 0.026   & 1.35 & \\ \midrule
\multirow{3}{*}{language}  & lang0          & 1.047   & 0.026  & 40.43 &$^{***}$\\
                           & lang1          & 0.569   & 0.036  & 15.79 &$^{***}$\\
                           & lang2          & 0.228  & 0.020  & 11.28  &$^{***}$\\ \midrule
\multirow{3}{*}{genre}     & genre0         & -0.023          & 0.022  & -1.05 & \\
                           & genre1         & 1.043  & 0.020  & 50.90 &$^{***}$\\
                           & genre2         & 0.061           & 0.041   & 1.51  & \\ \midrule
\multirow{3}{*}{mood}      & mood0          & 0.668   & 0.024  & 27.71  &$^{***}$ \\
                           & mood1          & 0.359  & 0.027  & 13.32 &$^{***}$ \\
                           & mood2          & 0.126  & 0.023   & 5.35  &$^{***}$ \\ \midrule
\multirow{2}{*}{diversity} & div0           & 0.770  & 0.015  & 50.59  &$^{***}$ \\
                           & div1           & 0.208  & 0.023   & 8.89 &$^{***}$ \\ \midrule
                           & Intercept      & 0.396  & 0.019  & 20.50 & $^{***}$ \\
\hline \hline
                           & Observations   & 112,753        \\
                           & $R^2$          & 0.1701         \\
                           & Adjusted $R^2$ & 0.1699         \\
                           & F-Statistic    & 1100 & & & $^{***}$   \\ \bottomrule
\end{tabular}
\end{table}

\subsection{Critical factors for homophily}
Given the rich backgrounds of active users carried by Weibo, we further investigate how characteristics of social ties, i.e., factors influence the homophily disclosed in music listening. Except for $sim_{song}$, other similarities referring to music preference are measured with $U_{c}$ (traits vectors of users), however, the vectors are also used to cluster the individuals. It could take a mistake when the dependent variable and some independent variables are calculated from the same source. Therefore, we only consider $sim_{song}$ as the dependent variable to predict in regression analysis. As can be seen, Table~\ref{tab:multiple_regression} presents the multiple regression results of nineteen factors function on the homophily.

As for gender, the coefficients of male are higher than that of female, indicating that relationships in online social network between female users are more homogeneous in music listening. Yet the correlations between listening similarities and the verification statuses are not significant due to trivial coefficients. For social attributes $NER0$ and $RFF0$ that reflecting individual extraversion and openness~\cite{Amichai2010Social,Bachrach2012Personality}, coefficients are significantly higher than that of $NER1$ and $RFF1$, implying that users of low extroversion and openness demonstrate more homophily in music listening.

For factors from perspective of music preferences, the positive coefficients shown in Table~\ref{tab:multiple_regression} indicate that users within the same group of clusterings referring to preferences of language and mood are similar. As for the language preference, users in the group of Chinese are more similar than those in other groups. Pop music and folk music are known as the mainstream genres in China. As for genres, it can be found that the users following pop and folk genre are significantly more similar. According to the results for languages and genres, we suggest that the friends in online social networks, whose music preferences accord with the mass taste, are always listening to the similar music. From the perspective of moods, the coefficients of the factor `exciting, wild and happy' is the highest for $sim_{song}$. It demonstrates that people who like the positive and energetic songs are more homogeneous in music listening. In addition, we investigate whether the diversity ($div0$ and $div1$) of music consumption is an influential factor for listening similarity. The results show that users with low musical diversity are more similar. However, users that enjoys multiple types of music, always own the uncommon and unique preferences and it is difficult for them to find friends sharing alike musical preferences.

\section{Discussion}
\label{sec:dis}

The booming of social media greatly facilitates exploitations of social networks by offering an unprecedented big-data window. In the present study, aiming at understanding homophily of music listening in online social networks, two networks with a same group of active users from Netease Music and Weibo are separately built for probing six measures of homophily. It is confirmed significantly that homophily exists in music listening of online social networks, even for the generic one from Weibo. Factors of social ties ranging from demographics to music preferences are further investigated in influencing homophily and many interesting patterns are revealed. To our best knowledge, for the first time a systematic and comprehensive study of music homophily is performed in a manner of data-driven solution. 

In the meantime, we are also the first to construct a big music data through matching active users in both music stream platform and Weibo. The unexpected existing of homophily in Weibo suggests that the knowledge from generic social networks can be confidently introduced and transfered to the study~\cite{malhotra2012studying} and application~\cite{carmagnola2009user} of domain-oriented social networks, which will essentially enrich the context of subjectives. Previous investigation shows that human averagely spend 17\% of life in music listening~\cite{rentfrow2012role}. Nevertheless, massive behavior data is still missing in existing studies. By combing behavioral and demographical data from different online social networks, our study demonstrates that explorations of music listening can be profoundly boosted.

In fact, music steaming platforms have been experiencing a fast growing stage with increasing influence through establishment of musical environments tailored to individual preferences. Plenty of music service providers use the relationships between online friends to estimate users' musical tastes and then recommend music~\cite{changtao2017systems}. In our study, we provide a solid evidence for the homophily of music listening both in specialized online social network and the generic network, suggesting that no matter specialized or generic social networks can serve as high quality sources of music recommendation~\cite{Sim2008Navigating,Bu2010Music}. We argue that the homophily should be exploited in building and enhancing music services. Specifically, the influential factors that function on listening similarities can be directly introduced into the design of music recommendation algorithms and coefficients we obtained can be principal weights of those factors. Taking gender as an example, music platforms could put more emphasis on female friends when recommend music to female users.

\section*{Acknowledgments}
This work was supported by NSFC (Grant No. 71501005) and the fund of the State Key Lab of Software Development Environment (Grant Nos. SKLSDE-2015ZX-05 and SKLSDE-2015ZX-28).

\section*{References}

\end{document}